\documentclass[smallextended]{svjour3}
\usepackage{graphicx}
\usepackage{amssymb,amsmath}

\newcommand{\Real}{\mathbb{R}}
\newcommand{\Complex}{\mathbb{C}}
\newcommand{\re}{\mbox{Re}}
\newcommand{\im}{\mbox{Im}}
\newcommand{\const}{\text{const}}

\begin{document}

\title{Quasi-normal acoustic oscillations in the transonic Bondi flow}

\author{Eliana Chaverra \and Olivier Sarbach}

\institute{Eliana Chaverra \at Instituto de F\'{\i}sica y Matem\'aticas,
Universidad Michoacana de San Nicol\'as de Hidalgo\\
Edificio C-3, Ciudad Universitaria, 58040 Morelia, Michoac\'an, M\'exico\\
\email{eliana@ifm.umich.mx}
\and Olivier Sarbach \at 
Instituto de F\'{\i}sica y Matem\'aticas,
Universidad Michoacana de San Nicol\'as de Hidalgo\\
Edificio C-3, Ciudad Universitaria, 58040 Morelia, Michoac\'an, M\'exico\\
and Perimeter Institute for Theoretical Physics, 31 Caroline St., Waterloo, ON, N2L 2Y5, Canada\\
\email{sarbach@ifm.umich.mx}}

\date{Received: date / Accepted: date}

\maketitle

\begin{abstract}
In recent work, we analyzed the dynamics of spherical and nonspherical acoustic perturbations of the Michel flow, describing the steady radial accretion of a relativistic perfect fluid into a nonrotating black hole. We showed that such perturbations undergo quasi-normal oscillations and computed the corresponding complex frequencies as a function of the black hole mass $M$ and the radius $r_c$ of the sonic horizon. It was found that when $r_c$ is much larger than the Schwarzschild radius $r_H = 2GM/c^2$ of the black hole, these frequencies scale like the surface gravity of the analogue black hole associated with the acoustic metric.

In this work, we analyze the Newtonian limit of the Michel solution and its acoustic perturbations. In this limit, the flow outside the sonic horizon reduces to the transonic Bondi flow, and the acoustic metric reduces to the one introduced by Unruh in the context of experimental black hole evaporation. We show that for the transonic Bondi flow, Unruh's acoustic metric describes an analogue black hole and compute the associated quasi-normal frequencies. We prove that they do indeed scale like the surface gravity of the acoustic black hole, thus providing an explanation for our previous results in the relativistic setting.

\keywords{accretion \and analogue black holes \and quasi-normal modes}
\end{abstract}

\section{Introduction}

Studying the accretion of matter into a black hole is a highly relevant subject in general relativity and astrophysics. From a purely theoretical point of view, it allows us to understand the dynamical evolution of black holes, a topic that has played a prominent role in the development of black hole thermodynamics~\cite{dC70,dCrR71,sH71,jBbCsH73} and is also directly related to the black hole stability problem~\cite{bKrW87,mDiR08}. On the astrophysical front, the study of accretion processes are utterly important since they provide a mechanism by which black holes can be ``observed". Traditionally, these observations are based on measuring the electromagnetic spectrum emitted by the accretion disk and comparing this spectrum to theoretical models (see Ref.~\cite{Shapiro-Book} and references therein). Recently, millimeter-wave very-long baseline interferometric arrays such as the Event Horizon Telescope~\cite{EHT} have been able to resolve the region around Sagittarius A$^*$, the supermassive black hole lying at the center of our galaxy, to scales smaller than its gravitational radius~\cite{sDetal08}. It is expected that the comparison between the observations to calculated images of the black hole shadow and the sharp photon ring surrounding it lead to tests for the validity of general relativity in its strong (albeit stationary) regime.

Another astrophysical scenario in which accretion comes into play is in a black hole-neutron star binary. For supermassive black holes, the star remains intact as it spirals around the black hole, and eventually it falls into it as a consequence of energy loss through gravitational radiation. However, for less massive black holes, more precisely, for black holes whose tidal radius is comparable or larger than the radius of its inner stable circular orbit, the star is tidally disrupted before it falls into the black hole, and an accretion disk forms. The formation of this disk has a significant impact on the gravitational wave signal, and so the study of this process is relevant for gravitational wave experiments such as LIGO~\cite{LIGO}, VIRGO~\cite{VIRGO} and KAGRA~\cite{KAGRA}, see~\cite{lLfP14} and references therein for a recent review on such binary systems.

In this article, we focus on the case of nearly spherical and steady-state accretion by a nonrotating black hole. Although this simple model does not say much about the scenarios described so far, it is still relevant in a number of astrophysical problems, like the accretion of interstellar matter by a stellar black hole~\cite{hB52,fM72,Shapiro-Book}, or the accretion of dark matter by a supermassive or stellar size black hole~\cite{fGfL11,fLmGfG14}. For recent work about radial accretion in a cosmological background, see also~\cite{jKeM13,pMeM13,pMeMjK13,pM15}.

In recent work~\cite{eCoS12,eCoS15a}, we provided a thorough analysis for the Michel flow~\cite{fM72}, describing the spherical, steady-state accretion into a nonrotating black hole. Unlike previous results which mainly focused on the case of a Schwarzschild background and a polytropic perfect fluid, our result in~\cite{eCoS15a} shows that given an arbitrary static, spherically symmetric black hole background satisfying certain reasonable conditions, and given a perfect fluid on this background with an arbitrary equation of state satisfying certain general conditions, there exists for each value $n_\infty$ of the particle density at infinity a unique accretion flow which is everywhere regular at and outside the event horizon. As it turns out, this flow is necessarily transonic, the fluid's velocity (as measured by static observers) being subsonic in the asymptotic region and supersonic close to the black hole.

In an extension of this work, in~\cite{eCmMoS15} we studied small, spherical and nonspherical acoustic perturbations of the Michel flow and showed that they exhibit quasi-normal acoustic oscillations. Their appearance is most easily understood by following the method by Moncrief~\cite{vM80} who showed that the dynamics of the acoustic perturbations are described in a very elegant manner by a wave equation on an effective curved background geometry determined by the \emph{acoustic (or sound) metric}. As it turns out, the geometry described by the acoustic metric has precisely the same features as a static and spherically symmetric black hole, where for the particular case of Michel accretion the horizon is located at the sonic sphere, and thus it describes a natural \emph{astrophysical analogue black hole}. In view of these observations, the appearance of quasi-normal acoustic oscillations is no surprise.

On the other hand, as discussed in~\cite{eCmMoS15}, the calculations of the corresponding modes and frequencies is technically challenging, since the acoustic metric is not known in explicit closed form. Consequently, traditional methods for computing the quasi-normal frequencies, such as Leaver's method~\cite{eL85} cannot easily be applied to this problem, and we had to design a new computation method to calculate the modes. Based on this new method, which is described in~\cite{eCmMoS15} and in more detail in~\cite{eCoS15b}, we computed the quasi-normal complex frequencies $s = \sigma + i\omega$ as a function of the black hole mass $M$ and the radius $r_c$ of the sonic sphere for a polytropic fluid with adiabatic index $\gamma = 4/3$. We noticed that for $r_c \gg r_H = 2GM/c^2$, both the decay rate $\sigma$ and the frequency $\omega$ of the oscillatory part of the mode scale like the surface gravity $\kappa$ of the acoustic black hole.

The main goal of this article is to provide an explanation for this behavior by reconsidering the accretion problem in the Newtonian limit. Here, by ``Newtonian limit" we mean the realistic case in which the speed of sound of the fluid $v_\infty$ at infinity is much smaller than the speed of light $c$. In this limit, as further discussed below, it can be shown that the accretion flow can be described to very good accuracy by the Newtonian equations in the region \emph{outside} (and including) the sonic horizon. The radial, steady state Newtonian solutions were studied in detail by Bondi~\cite{hB52} a long time ago, and as it turns out the Michel flow reduces to the transonic Bondi flow in the Newtonian limit~\cite{Shapiro-Book}.

After presenting a very brief, though self-contained derivation of the transonic Bondi flow solution, we derive the equations describing small acoustic perturbations thereof. Interestingly, the nonrelativistic description of this problem is still described by a wave equation with an effective curved metric background in four spacetime dimensions. The acoustic metric, which can be obtained from the Newtonian limit of Moncrief's sound metric, has been obtained by Unruh~\cite{wU81} in the context of an experimental black hole evaporation proposal. We discuss the geometry of the acoustic metric in the Newtonian case and compute the quasi-normal frequencies of the associated wave equation. In particular, we prove that these frequencies scale like the surface gravity of the acoustic hole and thereby provide an explanation for the results obtained in~\cite{eCmMoS15} in the relativistic case. Furthermore, we also discuss the eikonal approximation which connects the quasi-normal frequencies to the properties of the unstable circular null geodesics and find that it yields a fairly good description for the decay rates and the change in the oscillation frequencies as the angular momentum increases by one. Finally, we analyze the behavior of the quasi-normal frequencies $s$ with respect to the adiabatic index $\gamma$.

\section{Transonic Bondi flows}

The equations of motion describing a nonrelativistic, barotropic fluid in an external gravitational potential $\phi$, consist of the continuity equation
\begin{equation}
\dot{\rho} + \nabla\cdot (\rho{\bf v}) = 0
\label{Eq:Continuity}
\end{equation}
and the Euler equation
\begin{equation}
\dot{\bf v} + ({\bf v}\cdot\nabla){\bf v} + \nabla\left[ h(\rho) + \phi \right] = 0,
\label{Eq:Euler} 
\end{equation}
where $\rho$, ${\bf v}$ and $h(\rho) = \int dp/\rho$ refer to the density, the velocity and the enthalpy per unit mass of the fluid, respectively, and where a dot denotes partial differentiation with respect to time. For the following, we assume a polytropic equation of state, $p = K\rho^\gamma$, with $\gamma > 1$ the adiabatic index, in which case $h(\rho) = v_s^2/(\gamma-1)$, with $v_s = \sqrt{\partial p/\partial\rho} = \sqrt{K\gamma}\rho^{(\gamma-1)/2}$ the local speed of sound in the fluid.

For the purpose of the present article it is sufficient to consider the case where the fluid is irrotational, such that ${\bf v} = \nabla\psi$ with $\psi$ the fluid potential. In this case, Eq.~(\ref{Eq:Euler}) can be integrated
\begin{equation}
\dot{\psi} + \frac{1}{2} |{\bf v}|^2 + h(\rho) + \phi = \const.
\label{Eq:Bernoulli}
\end{equation}
This equation may be used to express $\rho$ in terms of first derivates of $\psi$, and when introduced into Eq.~(\ref{Eq:Continuity}) this yields a nonlinear wave equation for $\psi$. For a mathematical discussion regarding the formation of shocks based on the relativistic generalization of this wave equation, see~\cite{dC07,Christodoulou-Book}.

For the particular case of a radial fluid flow accreted by a central body of mass $M$, Eqs.~(\ref{Eq:Continuity},\ref{Eq:Euler}) can be integrated and give
\begin{eqnarray}
&& 4\pi r^2\rho u = \const,
\label{Eq:Bondi1}\\
&& \frac{1}{2} u^2 + \frac{v_s^2}{\gamma-1} - \frac{G M}{r} = \const,
\label{Eq:Bondi2}
\end{eqnarray}
where $u = v^r$ is the radial component of the velocity and $G$ is Newton's constant. The constant in the first equation is the mass accretion rate $\dot{M}$. Eqs.~(\ref{Eq:Bondi1},\ref{Eq:Bondi2}) describe an algebraic system of two equations for the three unknowns $(r,\rho,u)$. In order to analyze this system, following Refs.~\cite{hB52,Shapiro-Book} we introduce the dimensionless quantities
$$
\nu := \frac{u}{v_s},\quad
z := \frac{\rho}{\rho_\infty},\quad
x := \frac{r}{GM} v_\infty^2,\quad
\sigma := \frac{\gamma-1}{\gamma+1},\quad
\lambda := \frac{\dot{M}}{4\pi(GM)^2}\frac{v_\infty^3}{\rho_\infty}.
$$
where $\rho_\infty$ and $v_\infty$ refer to the density and sound speed, respectively, at infinity. With this notation Eqs.~(\ref{Eq:Bondi1}) can be rewritten as
\begin{equation}
x^2\nu z^{\frac{\gamma+1}{2}} = \lambda,
\label{Eq:ParticleCons}
\end{equation}
and Eq.~(\ref{Eq:Bondi2}) yields
\begin{equation}
f(\nu) = \lambda^{-2\sigma} g(x),
\label{Eq:BondiFlow}
\end{equation}
with the functions $f$ and $g$ defined by
\begin{eqnarray*}
f(\nu) &:=& \nu^{-2\sigma}\left( \frac{1}{2}\nu^2 + \frac{1}{\gamma-1} \right),\\
g(x) &:=& x^{4\sigma}\left( \frac{1}{x} + \frac{1}{\gamma-1} \right).
\end{eqnarray*}
%
%
%
The function $f : (0,\infty)\to \Real$ diverges to $+\infty$ as $\nu\to 0$ or $\nu\to\infty$ and has a global minimum at $\nu = \nu_c = 1$. Provided that $1 < \gamma < 5/3$, the function $g: (0,\infty)\to \Real$ also diverges to $+\infty$ as $x\to 0$ or $x\to \infty$, and in this case $g$ has a global minimum at $x = x_c = (5 - 3\gamma)/4$.

Although there are many other accretion flow solutions, in this article we are only interested in \emph{transonic} accretion flows, the flow being subsonic ($\nu < 1$) for large $x$ and supersonic ($\nu > 1$) close to the attractor. At the sonic sphere $\nu = \nu_c = 1$, where the flow changes from sub- to supersonic, the derivative of $f$ vanishes, and according to Eq.~(\ref{Eq:BondiFlow}) the derivative of $g$ also has to vanish there for the flow to be regular. Consequently, the transonic flow solution has to pass through the point $(x,\nu) = (x_c,\nu_c)$, and Eq.~(\ref{Eq:BondiFlow}) yields
\begin{equation}
\lambda = \lambda_c 
 = \frac{1}{4}\left( \frac{5 - 3\gamma}{2} \right)^{-\frac{5 - 3\gamma}{2(\gamma-1)}}.
\end{equation}
The compression rate at $x=x_c$ is
\begin{equation}
z_c = \left( \frac{5 - 3\gamma}{2} \right)^{-\frac{1}{\gamma-1}}.
\end{equation}
Outside the sonic sphere, the flow is uniquely determined by Eq.~(\ref{Eq:BondiFlow}) where $x$ and $\nu$ are restricted to the intervals $(x_c,\infty)$ and $(0,\nu_c)$, respectively, on which the functions $g$ and $f$ are monotonously increasing. Multiplying both sides of Eq.~(\ref{Eq:BondiFlow}) by $\nu^{2\sigma}$ and taking the limit $x\to \infty$, we find that along the flow
$$
x^2\nu \to \lambda_c
$$
in this limit, and consequently it follows from Eq.~(\ref{Eq:ParticleCons}) that $z\to 1$ and $v_s\to v_\infty$, as required. Therefore, the radial velocity of the flow, $u = v_s\nu$, converges to zero implying that the flow is at rest at infinity.

When $v_\infty$ is much smaller than the speed of light $c$, the transonic Bondi flow describes with very good accuracy the accretion flow into a Schwarzschild black hole in the region outside the sonic sphere. Indeed, when $v_\infty/c \ll 1$, and $1 < \gamma < 5/3$ the sonic sphere is located at a radius $r_c$ satisfying $r_c/r_H = (5-3\gamma)/(8v_\infty^2/c^2)$ (see for instance~\cite{eCoS15a}), with $r_H = 2GM/c^2$ the Schwarzschild radius of the hole, and hence the exterior of the sonic sphere lies entirely in the weak field region. Furthermore, since $u/c = (v_\infty/c) \nu z^{(\gamma-1)/2}$, the magnitude of the fluid's three-velocity, as measured by static observer, is much smaller than the speed of light in the region exterior to the sonic sphere. 

The relativistic generalizations of Eqs.~(\ref{Eq:Bondi1},\ref{Eq:Bondi2}) are (see, for instance~\cite{fM72,eCoS15a})
\begin{eqnarray}
4\pi r^2 n u^r &=& \const,
\label{Eq:jn}\\
h_{rel}(n)\sqrt{1 - \frac{r_H}{r} + \left( \frac{u^r}{c} \right)^2} &=& \const,
\label{Eq:je}
\end{eqnarray}
with $h_{rel}(n)$ the relativistic specific enthalpy and $n$ the particle density. In the nonrelativistic limit, $u^r\to u$, $n\to \rho/m$ with $m$ the particle mass, $h_{rel}(n) \to m[c^2 + h(\rho)]$, and using $h(\rho)\ll c^2$, $r_H/r\ll 1$, $u^r/c \ll 1$ for all $r \geq r_c$ one sees that Eqs.~(\ref{Eq:jn},\ref{Eq:je}) indeed reduce to Eqs.~(\ref{Eq:Bondi1},\ref{Eq:Bondi2}).

\section{Acoustic metric and the geometry it describes}

In this section, we consider spherical and nonspherical linearized acoustic perturbations of the Bondi flow, keeping the gravitational potential $\phi = -GM/r$ fixed. The dynamics of these perturbations can be obtained by replacing
$$
\rho \mapsto \rho + \varepsilon\delta\rho,\qquad
\psi \mapsto \psi + \varepsilon\delta\psi
$$
in Eqs.~(\ref{Eq:Continuity},\ref{Eq:Bernoulli}) and linearizing with respect to the small parameter $\varepsilon$, where $(\rho,{\bf v} = \nabla\psi)$ describes the (spherical and steady state) Bondi flow solution and $(\delta\rho,\delta{\bf v} = \nabla\delta\psi)$ the (non-spherical and time-dependent) acoustic perturbations. The resulting equations are
\begin{eqnarray}
&& \frac{\delta\dot{\rho}}{\rho} + ({\bf v}\cdot\nabla)\left( \frac{\delta\rho}{\rho} \right)
 + \left( \frac{\nabla\rho}{\rho} \right)\cdot\nabla\delta\psi  + \Delta\delta\psi = 0,\\
&& \delta\dot{\psi} + ({\bf v}\cdot\nabla)\delta\psi + v_s^2\frac{\delta\rho}{\rho} = 0,
\end{eqnarray}
where we have used the identity $\rho\partial h/\partial\rho = v_s^2$ and the background equation $\nabla\cdot(\rho{\bf v}) = 0$. Eliminating $\delta\rho/\rho$ from these two equations one obtains the following wave equation for $\delta\psi$ \cite{wU81}:
\begin{equation}
\Box_\mathfrak{G}\delta\psi = 0,
\label{Eq:AcousticWave}
\end{equation}
where the box operator $\Box_\mathfrak{G}$ is computed with respect to the \emph{acoustic metric}
\begin{equation}
\mathfrak{G} = \frac{\rho}{v_s}\left[ -v_s^2 dt^2 + \delta_{ij}(dx^i - v^i dt)(dx^j - v^j dt) \right].
\label{Eq:AcousticMetric}
\end{equation}
The fact that the dynamics of acoustic perturbations of an irrotational Newtonian fluid flow can be described by a wave equation on an effective curved background has been pointed out by Unruh~\cite{wU81} in the context of black hole analogue models and the possibility of measuring the analogue of Hawking radiation in these models. For recent experiments realizing such analogue models through fluid flows in the laboratory, see for example Ref.~\cite{sWeTmPwUgL11}.

Interestingly though, it seems that the description of the propagation of acoustic perturbations through an effective curved geometry had been found earlier by Moncrief~\cite{vM80} in an astrophysical context. In his work, Moncrief showed the propagation of acoustic linear perturbations of a relativistic, potential flow can be described by a wave equation on a curved background described by the sound metric. In the non-relativistic limit, Moncrief's sound metric reduces, up to a constant factor, to the acoustic metric given in Eq.~(\ref{Eq:AcousticMetric}). For the particular case of the Bondi flow we obtain
\begin{equation}
\mathfrak{G} = \frac{\rho}{v_s}\left[ -(v_s^2 - u^2) dt^2 - 2 u dt dr + dr^2 
 + r^2\left( d\vartheta^2 + \sin^2\vartheta d\varphi^2 \right) \right].
 \label{Eq:SoundMetricBondi}
\end{equation}
The acoustic metric~(\ref{Eq:SoundMetricBondi}) is spherically symmetric and possesses the Killing vector field
\begin{equation}
k = \frac{\partial}{\partial t}
\label{Eq:KVF}
\end{equation}
in the ``Newtonian'' spacetime $M = \Real\times \Real^3$ endowed with the metric $\mathfrak{G}$. In general relativity, the spacetime metric often provides a natural normalization for a timelike Killing vector field $k$. For example, if the spacetime is asymptotically flat, one may require that $k$ has unit norm at infinity. In the present case, there is no natural (fluid-independent) spacetime metric to provide the normalization; however we may normalize $k$ by requiring $k[t] = 1$ with $t$ the Newtonian time.

The negative square norm of the Killing vector field $k$ with respect to the acoustic metric is given by $N(r) := -\mathfrak{G}(k,k) = \rho v_s(1 - \nu^2)$. Consequently, $k$ is timelike in the subsonic region, spacelike in the supersonic region, and null at the sonic sphere $r = r_c$. Since the surface $r = r_c$ is a null surface, it constitutes a Killing horizon for the metric~(\ref{Eq:SoundMetricBondi}).\footnote{See for instance~\cite{Wald-Book,Heusler-Book} for a definition and properties of Killing horizons.} The associated ``surface gravity'' $\kappa$, defined by $\left. \nabla_\mu N \right|_{r=r_c} = \left. -2\kappa k_\mu \right|_{r=r_c}$, will play an important role later. It is given by
\begin{equation}
\kappa = -\left. v_s\frac{d\nu}{dr} \right|_{r=r_c} = \frac{v_\infty}{r_c}
 = \frac{4}{5-3\gamma}\frac{v_\infty^3}{GM} .
\label{Eq:SurfaceGravity}
\end{equation}
In deriving this result, we have differentiated Eq.~(\ref{Eq:BondiFlow}) twice and evaluated the result at $x = x_c$ in order to compute the first derivative of $\nu$ at the critical point, which yields
\begin{equation}
\left. \frac{d\nu}{dx} \right|_{x_c} = -\sqrt{\frac{8}{5-3\gamma}}.
\label{Eq:dnudxc}
\end{equation}
Notice that with our normalization of the Killing vector field and definition of the surface gravity, $\kappa$ has units of frequency. This should be contrasted to the case of static or stationary black hole spacetimes in general relativity, where due to a different normalization involving the speed of light, the surface gravity is interpreted as an acceleration, see~\cite{Wald-Book}. For the purpose of the present article and the calculation below of the quasi-normal acoustic frequencies, the expression~(\ref{Eq:SurfaceGravity}) not only has the adequate units, but in fact we will see that it also provided the correct scaling for the frequencies.

For the following computations it is convenient to rewrite the acoustic metric~(\ref{Eq:SoundMetricBondi}) in diagonal form by means of a suitable coordinate transformation. Introducing the new time coordinate~\cite{wU81},
$$
T = t + \int\frac{u dr}{v_s^2 - u^2},
$$
the metric outside the sonic horizon assumes the form:
\begin{equation}
\mathfrak{G} = \frac{\rho}{v_s}\left[ -(1-\nu^2) v_s^2 dT^2 + \frac{dr^2}{1 - \nu^2}
 + r^2\left( d\vartheta^2 + \sin^2\vartheta d\varphi^2 \right) \right].
\label{Eq:SoundMetricBondiDiag}
\end{equation}
For large radii, $x \gg \gamma-1$, we have $x^2\nu \simeq \lambda$ and $\rho\simeq \rho_\infty$, $v_s\simeq v_\infty$, and hence the acoustic metric becomes (up to a constant conformal factor)
\begin{equation}
\mathfrak{G} \simeq -\left( 1 - \frac{\lambda^2}{x^4} \right) (v_\infty dT)^2 
 +  \frac{dr^2}{1 - \frac{\lambda^2}{x^4}} 
  + r^2\left( d\vartheta^2 + \sin^2\vartheta d\varphi^2 \right),
\label{Eq:SoundMetricBondiAsymptotic}
\end{equation}
which shows, in particular, that the acoustic metric is asymptotically flat.\footnote{Note that the asymptotic metric~(\ref{Eq:SoundMetricBondiAsymptotic}) is equal to the ``canonical acoustic hole" introduced in~\cite{mV98}. Naively, one might think that this metric describes the limit $\gamma\to 1$ of the metric~(\ref{Eq:SoundMetricBondiDiag}); however as we will discuss further below the relation $x^2\nu = \lambda$ used in deriving the asymptotic form~(\ref{Eq:SoundMetricBondiAsymptotic}) of the metric does not hold for arbitrary values of $x$ in the limit $\gamma\to 1$.}

Summarizing this section, we have found that in what the propagation of sound waves is concerned, the geometry described by the metric~(\ref{Eq:SoundMetricBondi}) has the same qualitative features as a static, spherically symmetric and asymptotically flat black hole. The sonic horizon $r = r_c$ plays the role of the event horizon in this analogue acoustic black hole.

\section{Quasi-normal acoustic frequencies}

In the previous section, we discussed the geometric properties of the acoustic metric for the transonic Bondi flow. Since this metric describes an acoustic hole, and since the acoustic perturbations of the transonic Bondi flow are governed by a wave equation on the effective geometry described by this metric, it is not surprising to find that when perturbed, this flow exhibits quasi-normal acoustic oscillations. In the relativistic case, we showed that such oscillations arise from fairly arbitrary initial excitations~\cite{eCmMoS15} by solving numerically the Cauchy problem for the wave equation~(\ref{Eq:AcousticWave}). Here, we do not compute the full solution of the wave equation, but rather we focus on the quasi-normal mode solutions of Eq.~(\ref{Eq:AcousticWave}) and compute the associated complex frequencies $s = \sigma + i\omega\in \Complex$ based on our matching method described in~\cite{eCmMoS15}.

The quasi-normal modes have the form
\begin{equation}
\delta\psi = \frac{1}{R} e^{s T}\phi(s,r) Y^{\ell m}(\vartheta,\varphi),
\label{Eq:deltapsi}
\end{equation}
where $R := \sqrt{\rho/v_s}\, r$ the areal radius associated with the acoustic metric, $\phi(s,r)$ is a complex-valued function to be determined, and $Y^{\ell m}$ denote the standard spherical harmonics with angular momentum number $\ell m$. Introducing the ansatz~(\ref{Eq:deltapsi}) into Eq.~(\ref{Eq:AcousticWave}) and using the diagonal parametrization~(\ref{Eq:SoundMetricBondiDiag}) of the acoustic metric, we obtain the following ``eigenvalue" problem:
\begin{equation}
 - {\cal N}(r)\frac{\partial}{\partial r} \left[ {\cal N}(r)\frac{\partial \phi}{\partial r} \right]
 + \left[ s^2 + {\cal N}(r) V_\ell(r)\right]\phi = 0,
\label{Eq:phi}
\end{equation}
with the function ${\cal N}$ and the effective potential $V_\ell$ given by
\begin{equation}
{\cal N}(r) = v_s(1 - \nu^2),\qquad
V_\ell(r) =  v_s\frac{\ell(\ell+1)}{r^2} 
 + \frac{1}{R}\frac{d}{dr}\left( {\cal N}(r)\frac{dR}{dr} \right).
\end{equation}
The quasi-normal frequencies are determined by the following requirement on $\phi$~\cite{hNbS92}: Let $\phi_+(s,r)$ and $\phi_-(s,r)$ be the unique solutions of Eq.~(\ref{Eq:phi}) satisfying the boundary conditions
\begin{equation}
\lim\limits_{r_*\to\infty} e^{s r_*}\phi_+(s,r) = 1,\quad
\lim\limits_{r_*\to -\infty} e^{-s r_*}\phi_-(s,r) = 1,\quad
r_* := \int\limits^r\frac{dr'}{{\cal N}(r')},
\label{Eq:phiBC}
\end{equation}
for each $\re(s) > 0$, and consider their analytic continuation to the left complex plane $\re(s) < 0$. Then, the quasi-normal frequencies are determined by those values of $s$ for which the functions $\phi_+(s,\cdot)$ and $\phi_-(s,\cdot)$ are linearly dependent. In~\cite{eCmMoS15} we discussed a new numerical method for computing the analytic continuations of the functions $\phi_\pm(s,\cdot)$ based on an iteration scheme, where each iteration involves computing numerically an integral over a suitable path in the complex $r$-plane. Based on this method, the quasi-normal frequencies can be found by determining the zeros of the Wronskian
$$
W(s) := \det\left( \begin{array}{rr}
\phi_+(s,r) & \phi_-(s,r)\\
\frac{\partial\phi_+}{\partial r_*}(s,r) & \frac{\partial\phi_-}{\partial r_*}(s,r)
\end{array} \right)
$$
based on a standard Newton algorithm~\cite{Recipes-Book}.

In terms of dimensionless quantities, Eq.~(\ref{Eq:phi}) can be rewritten as
\begin{equation}
 - \overline{\cal N}(x)\frac{\partial}{\partial x}
  \left[ \overline{\cal N}(x)\frac{\partial \phi}{\partial x} \right]
 + \left[ \overline{s}^2 + \overline{\cal N}(x) \overline{V_\ell}(x)\right]\phi = 0,
\label{Eq:phiBis}
\end{equation}
with
\begin{eqnarray}
\overline{\cal N}(x) &=& \frac{1 - \nu^2}{(x^2\nu)^\sigma},
\label{Eq:Nbar}\\ 
\overline{V_\ell}(x) &=& \frac{1}{(x^2\nu)^\sigma}\frac{1}{x^2} \left[ 
 \ell(\ell+1) - 2\sigma(1-\nu^2) - \sigma(1 + 3\nu^2)\frac{x\nu_x}{\nu}
 \right. \nonumber\\
 && \left. \qquad
  - \frac{\gamma-3}{4(\gamma+1)}(3 + \nu^2)\left( \frac{x\nu_x}{\nu} \right)^2
  + \frac{\gamma-3}{2(\gamma+1)}(1-\nu^2)\frac{x^2\nu_{xx}}{\nu} \right].
\label{Eq:Vlbar}
\end{eqnarray}
Here an in the following, $\nu_x$ and $\nu_{xx}$ denote the first and second derivatives of $\nu$ with respect to $x$. Since asymptotically $\nu \simeq \lambda/x^2$, the effective potential $\overline{V_\ell}$ decays as fast $1/x^2$ when $x\to \infty$. In fact, it follows that $\lim_{x\to\infty} x^2\overline{V_\ell}(x) = \ell(\ell+1)/\lambda^{2\sigma}$, which shows that the centrifugal term dominates for large $x$. The first derivative $\nu_x$ can be computed from
$$
\frac{x\nu_x}{\nu} = \lambda^{-2\sigma}\frac{x g_x(x)}{\nu f_\nu(\nu)},
$$
as long as $x\neq x_c$, whereas for $x = x_c$ one can use the expression~(\ref{Eq:dnudxc}). The second derivative $\nu_{xx}$ can be computed using the identity
$$
\frac{2}{\gamma+1}(1-\nu^2)\frac{x^2\nu_{xx}}{\nu} = -\nu^{2\sigma}
\left( \lambda^{-2\sigma} x^2 g_{xx}(x) - f_{\nu\nu}(\nu) x^2\nu_x^2 \right),
$$
which follows from differentiating Eq.~(\ref{Eq:BondiFlow}) twice, and this also shows that the effective potential is regular at the sonic horizon. In Eq.~(\ref{Eq:phiBis}) the dimensionless frequency $\overline{s}$ is related to $s$ via the relation
$$
s = \lambda^\sigma\frac{v_\infty^3}{GM}\overline{s},
$$
which shows that $s$ must scale like the surface gravity $\kappa$ of the acoustic black hole, see Eq.~(\ref{Eq:SurfaceGravity}). 

For the computation of the functions $\overline{\cal N}(x)$ and $\overline{V_\ell}(x)$, one needs to determine $\nu$ as a function of $x$, which has to be determined by inversion of the function $f$ in Eq.~(\ref{Eq:BondiFlow}). We invert $f$ numerically, using a standard Newton algorithm~\cite{Recipes-Book}, starting from the seed value
$$
\nu_0(x) = \frac{1}{[1 + B_+(x-x_c)][1 + B_-(x-x_x)]},\qquad
B_\pm := \frac{1 \pm \sqrt{1 - \frac{5-3\gamma}{2\lambda}} }{\sqrt{\frac{5-3\gamma}{2}}},
$$
which has the correct asymptotic behavior for small $|x-x_c|$ (see Eq.~(\ref{Eq:dnudxc})) and for large $x$, for which $\nu\simeq \lambda/x^2$.

Results for the frequencies of the fundamental quasi-normal acoustic oscillations of a polytropic fluid with adiabatic index $\gamma = 1.3333 \simeq 4/3$ and different values for the angular momentum number $\ell$ are shown in Table~\ref{Tab:QNM}. Note that for high $\ell$ the real part of $s$ seems to converge to a value close to $-0.39\kappa$ while the imaginary part augments by a value close to $0.6\kappa$ as $\ell$ increases by one. As we will discuss in the next section, these features can be understood from simple analytic considerations in the high-frequency limit.
\begin{table}[h]
\center
\begin{tabular}{|r|l|l|}\hline
$\ell$ & $s/\kappa$ & $\Delta(\im(s)/\kappa)$ \\
\hline
$1$ & $ -0.3863 +  0.8217i$ &
\\ 
$2$ &$  -0.3887 +  1.455i$ & 0.633
\\
$3$ &  $-0.3893 +  2.069i$ & 0.614
\\
$4$ & $-0.3896+  2.678i$ & 0.609
\\
$5$ & $ -0.3898+  3.284i$ & 0.606
\\ 
$6$ &   $-0.3898 + 3.888i$ & 0.602
\\
$7$ &  $-0.3899 + 4.491i$ & 0.603
\\
$8$ & $-0.3899 + 5.094i$ & 0.603
\\
$9$ & $-0.3899 +  5.697i$ & 0.603
\\
$10$ &  $-0.3899 + 6.299i $ & 0.602 
\\
\hline
\end{tabular}
\caption{Fundamental quasi-normal frequencies for acoustic perturbations of a Bondi flow with adiabatic index $\gamma = 1.3333$ and angular momentum number $\ell$. Four significant figures are shown. The last column shows the difference between the imaginary part of $s/\kappa$ for the given value of $\ell$ and the same quantity for the previous value of $\ell$.}
\label{Tab:QNM}
\end{table}

\begin{figure}[htp]
\begin{center}
\includegraphics[width=11.0cm]{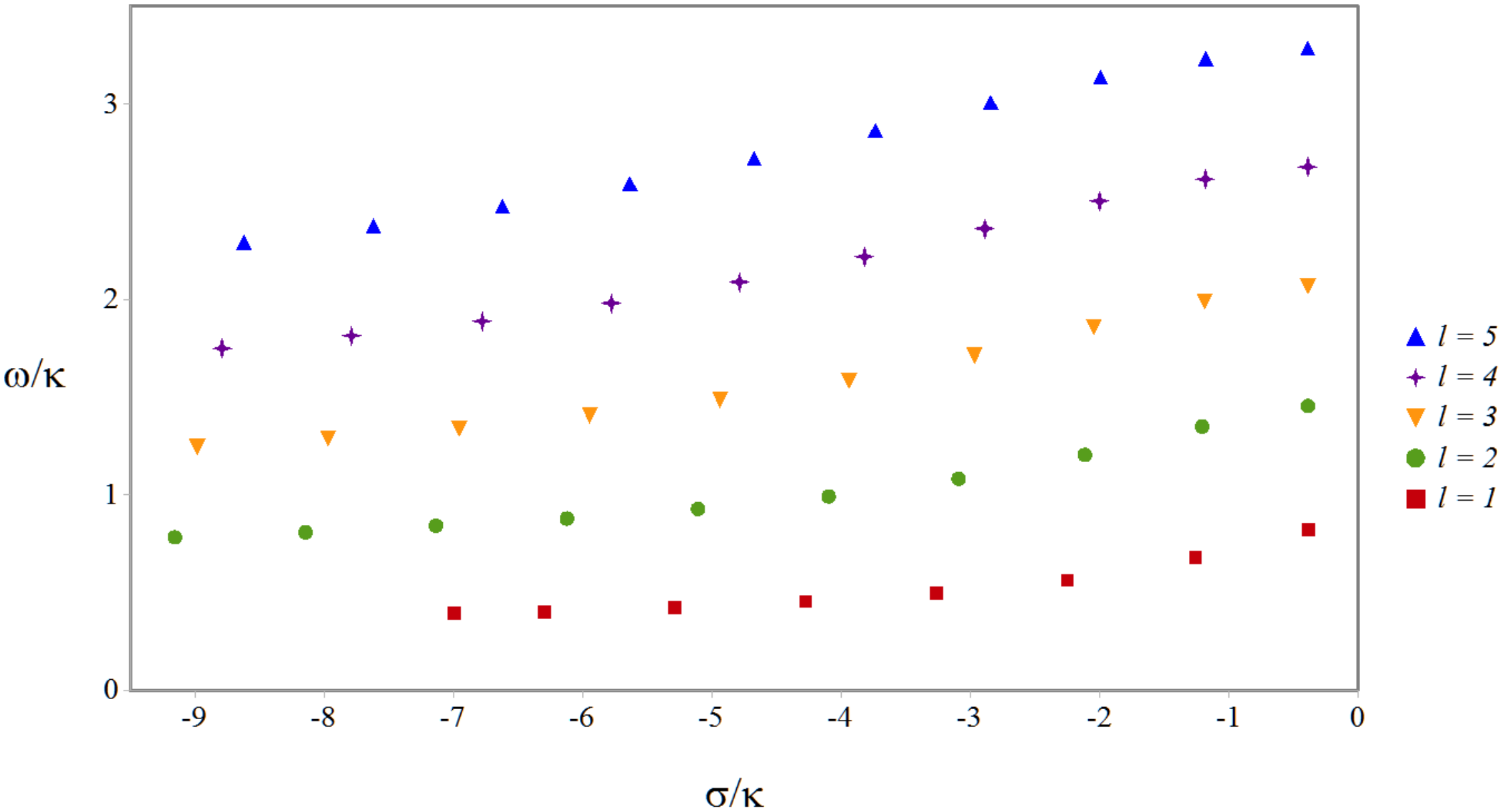}
\end{center}
\caption{The spectrum of the quasi-normal acoustic frequencies of a polytropic Bondi flow with adiabatic index $\gamma = 1.3333$.}
\label{Fig:Overtones}
\end{figure}

The frequencies for the first few overtones are given in Table~\ref{Tab:QNMExcited}, and the corresponding spectrum is shown in Fig.~\ref{Fig:Overtones}.
\begin{table}[h]
\center
\begin{tabular}{| r | c | c | c | c | c |}
\hline
$n_o$ & $s/\kappa$ $(\ell = 1)$ & $s/\kappa$ $(\ell = 2)$ 
 & $s/\kappa$ $(\ell = 3)$ & $s/\kappa$ $(\ell = 4)$ & $s/\kappa$ $(\ell = 5)$
\\
\hline
$1$ & $-1.259 + 0.6782i$ & $-1.207 + 1.350i$ & $-1.189 + 1.991i$ & $ -1.182 + 2.615i $  & $-1.178 +3.231i $ \\
$2$ & $-2.249 + 0.5629i$ & $-2.115 + 1.204i$ & $-2.045 + 1.859i$ & $-2.001 +2.503i $    & $ -1.991+ 3.136i$\\  
$3$ & $-3.262 + 0.4965i$ & $-3.091 + 1.082i$ & $-2.967 + 1.714i$ & $-2.888 + 2.362i$    & $-2.841 + 3.007i$\\
$4$ & $-4.275 + 0.4539i$ & $-4.097 + 0.9924i$ & $-3.938 + 1.587i$ & $-3.819 + 2.219i$  & $-3.735 + 2.865i$\\
$5$ & $-5.287 + 0.4237i$ & $-5.108 + 0.9279i$ & $-4.935 + 1.486i$ & $-4.787+ 2.089i$   & $-4.670 + 2.722i$\\
$6$ & $-6.295 + 0.4007i$ & $-6.122 + 0.8789i$ & $-5.943 + 1.406i $ &$-5.777 + 1.981i$ & $-5.634 + 2.591i$\\
$7$ & $-6.993 + 0.3942i$  & $-7.135 + 0.8422i$ & $-6.955 +1.342i$  &$-6.778 + 1.889i$   & $-6.619 +2.477i$\\
$8$ &                                    & $-8.144 + 0.8087i$ & $-7.968 + 1.289i $ & $-7.787 + 1.814i$  & $-7.618 +2.378i$ \\
$9$ &                                    & $-9.155 + 0.7825i $ & $-8.980+ 1.246i$ & $-8.798 + 1.749i$ & $-8.622 + 2.292i$ \\
\hline
\end{tabular}
\caption{First few overtones of the acoustic frequencies shown in the previous table. Here, $n_o$ stands for the overtone number, $n_o = 1$ denoting the first overtone, $n_o = 2$ the second one, etc.}  
\label{Tab:QNMExcited}
\end{table}

\section{Eikonal limit}

\begin{figure}[htp]
\begin{center}
\includegraphics[width=7.0cm]{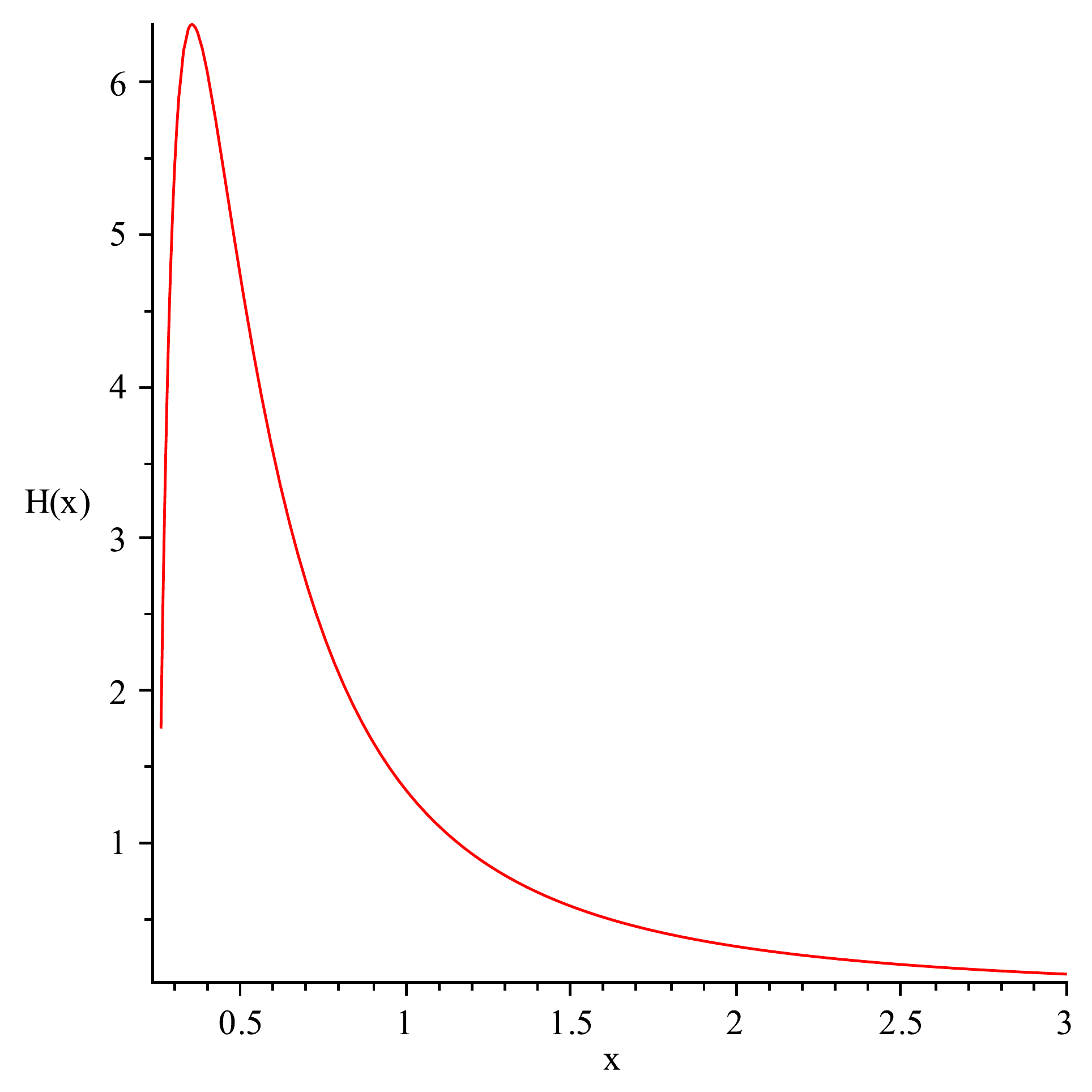}
\end{center}
\caption{A plot of the function $\overline{H}(x)$ for $\gamma = 4/3$. There is a maximum at $x = x_{circ} \simeq 0.35396$ where $\nu = \nu_{circ}\simeq 0.76842$.}
\label{Fig:FunctionH}
\end{figure}

As we show in this section, the quasi-normal frequencies found numerically in the previous section can be approximated to a fairly good accuracy using the high-frequency limit, in which case the quasi-normal frequencies can be obtained directly from the properties of unstable circular null geodesics. As discussed in~\cite{vCaMeBhWvZ09,eBvCaS09} and references therein, in the high-frequency limit the quasi-normal oscillations can be interpreted in terms of decaying wave packets which are localized along an unstable circular null geodesic. Denoting by $\Omega_{circ}$ and $\Lambda_{circ}$ the angular velocity and the Lyapunov exponent, respectively, of the unstable null geodesic, the quasi-normal frequencies for high values of $\ell$ are given by the formula~\cite{vCaMeBhWvZ09} 
\begin{equation}
s = -\left( n_o + \frac{1}{2} \right)\Lambda_{circ} + i\ell\,\Omega_{circ},
\label{Eq:QNMEikonalSpectrum}
\end{equation}
with $n_o$ the overtone number. For an arbitrary asymptotically flat, static spherically symmetric metric of the form
\begin{equation}
ds^2 = -F(r) dT^2 + \frac{dr^2}{G(r)} 
 + r^2\left( d\vartheta^2 + \sin^2\vartheta d\varphi^2 \right)
\label{Eq:SphSymMetric}
\end{equation}
with time coordinate $T$ and positive smooth functions $F(r)$ and $G(r)$, there exists an unstable circular null geodesic at $r = r_{circ}$ if and only if the function $H(r) := F(r)/r^2$ has a local maximum at $r = r_{circ}$, and in this case the associated angular velocity and Lyapunov exponent are given by~\cite{vCaMeBhWvZ09}
\begin{equation}
\Omega_{circ} = \sqrt{H(r_{circ})},\quad
\Lambda_{circ} = \left. \sqrt{\frac{F(r) G(r)}{2}}
 \sqrt{ -\frac{1}{H(r)} \frac{d^2}{dr^2} H(r) } \right|_{r = r_{circ}}.
\end{equation}
Comparing Eq.~(\ref{Eq:SphSymMetric}) with Eq.~(\ref{Eq:SoundMetricBondiDiag}) and discarding the conformal factor $\rho/v_s$ which does not affect the null geodesics as trajectories in spacetime we find $H(r) = v_s^2(1-\nu^2)/r^2$, $\sqrt{F(r) G(r)} = v_s(1-\nu^2)$, such that
\begin{eqnarray}
\Omega_{circ} &=& \lambda^\sigma\frac{v_\infty^3}{GM}\sqrt{\overline{H}(x_{circ})},
\\
\Lambda_{circ} &=& \lambda^\sigma\frac{v_\infty^3}{GM}
\left. \frac{1-\nu^2}{(x^2\nu)^\sigma}
 \sqrt{ -\frac{1}{2\overline{H}(x)} \frac{d^2}{dx^2} \overline{H}(x) } \right|_{x = x_{circ}},
\end{eqnarray}
with the dimensionless function
$$
\overline{H}(x) := \frac{1-\nu^2}{x^{4\sigma+2}\nu^{2\sigma}},
$$
where $\nu$ is implicitly determined by Eq.~(\ref{Eq:BondiFlow}). A plot of $\overline{H}(x)$ for the typical value $\gamma = 4/3$ is shown in Fig.~\ref{Fig:FunctionH}, and as is visible from this plot there is a unique unstable circular null geodesic corresponding to the maximum of this function. 
Computing this maximum numerically, we find, for $\gamma = 4/3$,
\begin{equation}
\frac{\Lambda_{circ}}{2\kappa} \simeq 0.39006,\qquad
\frac{\Omega_{circ}}{\kappa} \simeq 0.60095,
\end{equation}
which agrees very well with the corresponding values in Table~\ref{Tab:QNM}. In fact, the formula~(\ref{Eq:QNMEikonalSpectrum}) also fits reasonably well with the first few overtones in Table~\ref{Tab:QNMExcited}. For example, for $n_o = 3$ Eq.~(\ref{Eq:QNMEikonalSpectrum}) predicts $\re(s)/\kappa\simeq -2.73042$ which agrees with the corresponding value for $\ell = 5$ in Table~\ref{Tab:QNMExcited} to about $4\%$ accuracy. For higher values of $n_o$, a good fit seems to require higher values of $\ell$.

\section{Dependency on $\gamma$}

Finally, we analyze the dependency of the fundamental frequencies on the adiabatic index $\gamma$. Table~\ref{Tab:DependGammaSigma} shows such frequencies for $\ell=1,2,3$ and four different values of $\gamma$, varying in the interval $1 < \gamma < 5/3$. The results from the eikonal approximation described in the previous section are also given for comparison. As for the case $\gamma = 4/3$, the Lyapunov exponent provides a very accurate description for the decay rate, and the angular velocity gives a very good description for the increment in the oscillation frequency as $\ell$ increases by one.
\begin{table}[h]
\center
\begin{tabular}{| c | c | c | c | c | l |}
\hline
$\gamma$ &  $s/\kappa$ ($\ell = 1$) & $s/\kappa$ ($\ell = 2$) 
 & $s/\kappa$ ($\ell=3$)  &  $\Lambda_{circ}/(2\kappa)$ &  $\Omega_{circ}/\kappa$
\\
\hline
$1.66$ & $-0.4345 + 2.260i$ & $-0.4343 + 3.924i$ & $-0.4343 + 5.554i$ 
& $  0.4343              $  & $ 1.604 $ \\
$1.50$ & $-0.4048 + 0.9858i$ & $-0.4053 + 1.732i$ & $-0.4055 + 2.459i$ 
& $0.4057 $    & $ 0.7125$\\  
$1.15$ & $-0.3698 + 0.7335i$ & $-0.3743 + 1.307i$ & $-0.3755 + 1.861i$ 
& $0.3768$    & $0.5414$\\
$1.01$ & $-0.3591+ 0.6905i$ & $-0.3650 + 1.234i$ & $-0.3666 + 1.759i$ 
& $0.3681$  & $0.5121$\\
\hline
$1.00$ & $-0.3584 + 0.6879i$ & $-0.3644 + 1.229i$ & $-0.3660 + 1.753i$ & & \\
\hline
\end{tabular}
\caption{Fundamental quasi-normal acoustic frequencies for different values of $\gamma$. Four significant figures are shown. The values for $\gamma = 1.00$ are obtained using the limiting procedure described below.}
\label{Tab:DependGammaSigma}
\end{table}

\begin{figure}[htp]
\begin{center}
\includegraphics[width=5.6cm]{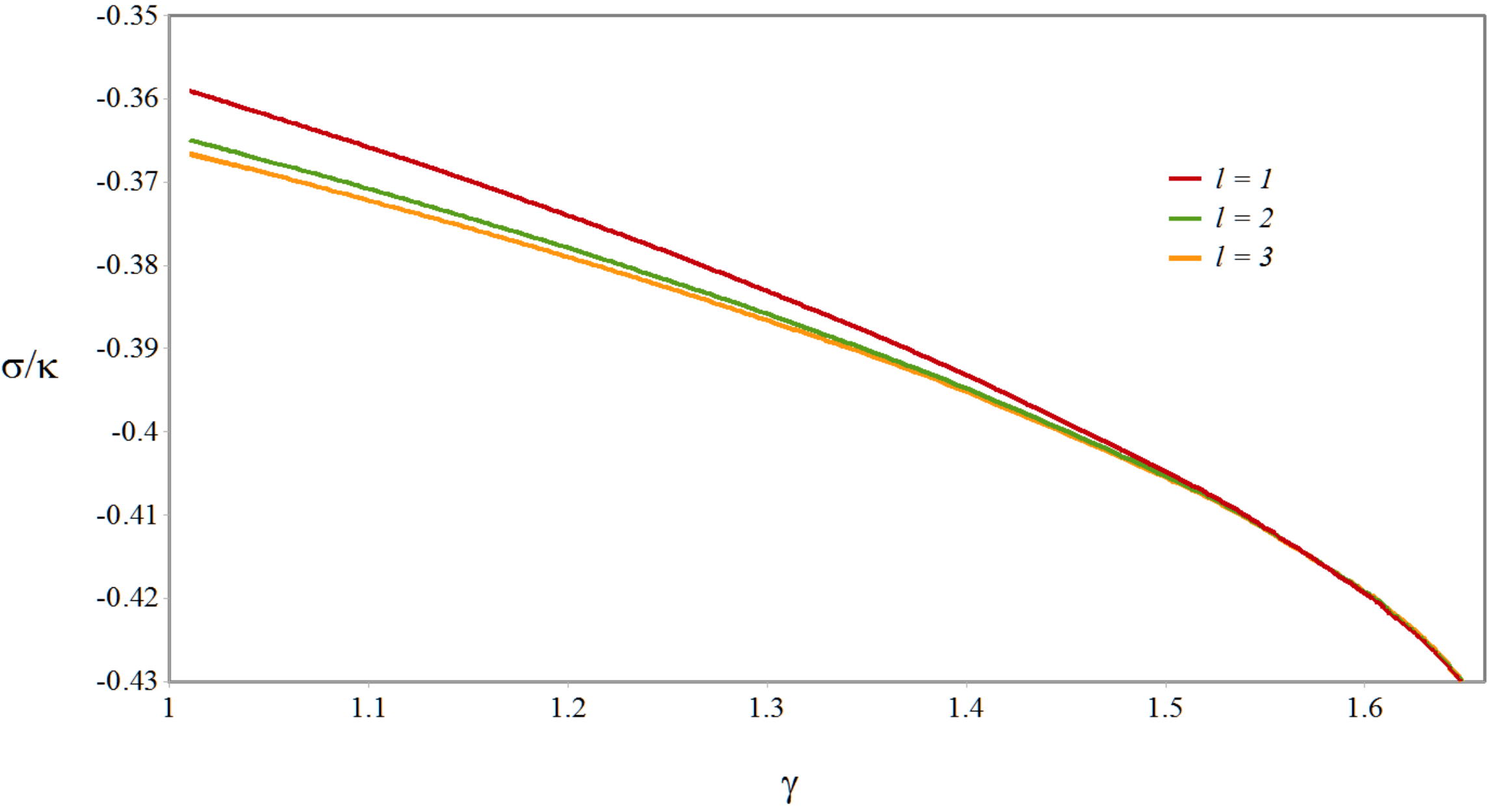}
\includegraphics[width=5.6cm]{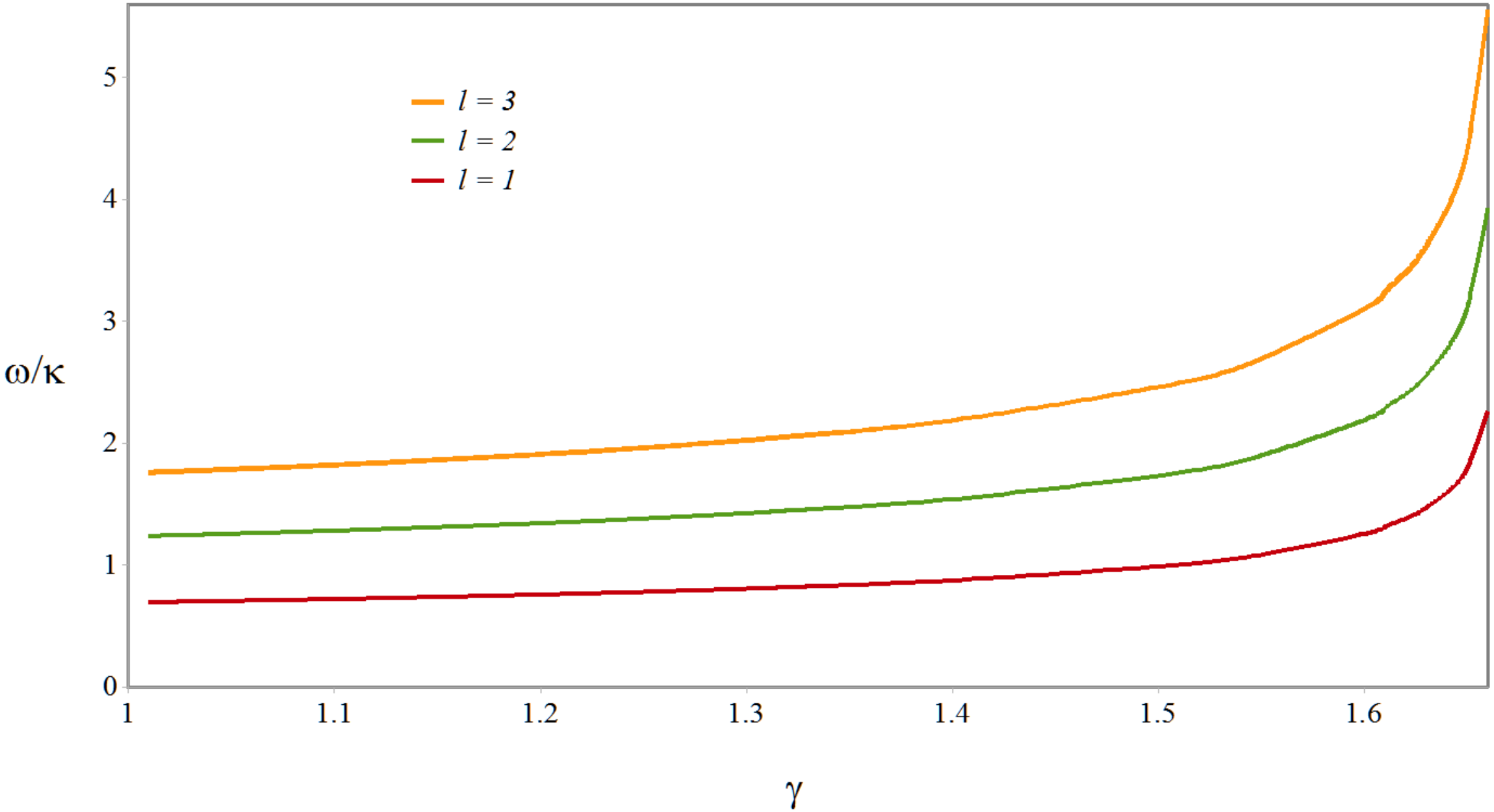}
\end{center}
\caption{Fundamental quasi-normal acoustic frequency as a function of the adiabatic index $\gamma$ in the range $1 < \gamma < 5/3$. Left panel: real part of $s$. Right panel: imaginary part of $s$.}
\label{Fig:DependGammaSigma}
\end{figure}

In Fig.~\ref{Fig:DependGammaSigma} we provide plots for the dependency of the decay rate and oscillation frequencies on $\gamma$. Empirically, we found that
$$
\frac{\omega}{\kappa} \simeq C_\ell \left( \frac{5}{3} - \gamma \right)^{-\beta},\qquad
\ell = 1,2,3,
$$
with $C_\ell$ a constant depending on $\ell$ and $\beta$ a constant which is approximately equal to $0.25$. In particular, this implies that $\omega/\kappa$ diverges as $\gamma\to 5/3$. This makes sense since (at the Newtonian level considered here) there is no transonic, steady-state radial accretion flow when $\gamma\geq 5/3$.

In contrast to this, $\sigma/\kappa$ and $\omega/\kappa$ have finite limits when $\gamma\to 1$. This limit can be understood by first rewriting Eq.~(\ref{Eq:BondiFlow}) in the form
$$
\frac{x^2\nu}{\lambda} 
 = \left( \frac{1 + \frac{1}{2}(\gamma-1)\nu^2}{1 + \frac{\gamma-1}{x}} \right)^{1/2\sigma}
$$
and taking the limit $\gamma\to 1$ on both sides, keeping $x$ fixed. Noting that $\lambda\to e^{3/2}/4$, and assuming that $\nu$ converges to a finite value in this limit, we obtain the implicit relation
$$
x^2\nu = \frac{1}{4} e^{(\nu^2+3)/2 - 1/x}
$$
between $x$ and $\nu$. As a consequence, $(x^2\nu)^\sigma\to 1$, and the functions $\overline{\cal N}$ and $\overline{V_\ell}$ in Eqs.~(\ref{Eq:Nbar},\ref{Eq:Vlbar}) simplify to
\begin{eqnarray*}
\overline{\cal N}(x) &=& 1 - \nu^2,\\
\overline{V_\ell}(x) &=& \frac{\ell(\ell+1)}{x^2} + \frac{1}{x^2}\left[ \frac{(1-\frac{1}{2x})^2}{1-\nu^2} + \frac{1}{x} -1  \right].
\end{eqnarray*}
Computing the fundamental quasi-normal frequencies directly with these expressions yields the results shown in the last line of Table~\ref{Tab:DependGammaSigma}, which provide the correct limiting values as $\gamma\to 1$.

\section{Conclusions}

We have analyzed linear acoustic perturbations of the Michel flow solution, describing a transonic radial accretion flow into a nonrotating black hole. Whereas previous work by the authors~\cite{eCmMoS15} treated this scenario using the full relativistic Euler equations, in the present article we focused on the realistic case where the sound speed at infinity is much smaller than the speed of light. In this limit, the accretion flow at and outside the sonic horizon is accurately described by the nonrelativistic Euler equations, and thus a Newtonian description is sufficient for an understanding of the behavior of the flow outside the sonic horizon. The solutions for the unperturbed flow in this case have been given by Bondi~\cite{hB52}, and the dynamics of its acoustic perturbations are described in an elegant and efficient way by a wave equation on an effective curved background geometry described by the acoustic metric introduced by Unruh~\cite{wU81}. For the transonic Bondi flow, the causal structure described by the acoustic metric has the same properties as a static, spherically symmetric black hole whose event horizon is described by the sonic horizon. Therefore, as far as the propagation of sound waves is concerned, the acoustic geometry has the same qualitative properties as a black hole, and thus it describes an analogue black hole~\cite{cBsLmV05}.

In analogy with a scalar field propagating on a stationary black hole background, the acoustic perturbations of the Bondi flow exhibit quasi-normal oscillations. Using the analogue black hole description, we have computed the associated quasi-normal frequencies $s = \sigma + i\omega$, where $\sigma < 0$ describes the decay rate and $\omega$ the oscillation frequency. In the fully relativistic case, we had computed these frequencies numerically~\cite{eCmMoS15} and found empirically that they scaled like the surface gravity $\kappa$ of the acoustic black hole in the limit where the sonic horizon is much larger than the Schwarzschild radius of the black hole. In the present article we have managed to provide a proof for this behavior by reconsidering the problem in the Newtonian limit and by showing explicitly that $s$ scales like $\kappa$. In particular, our Newtonian calculations show that for high values of the angular momentum number $\ell$, the fundamental quasi-normal acoustic frequencies of a polytropic fluid with adiabatic index $\gamma = 1.3333 \simeq 4/3$, the real part of $s$ converges to a value close to $-0.39\kappa$ while the imaginary part augments by a value close to $0.6\kappa$ as $\ell$ increases by one. As we further showed in this article, these features can be understood from an analysis in the high-frequency limit. Therefore, our results provide an explanation for the empirical formula~(45) given in Ref.~\cite{eCmMoS15}.

We have also analyzed the dependency of the quasi-normal acoustic frequencies on the adiabatic index $\gamma$ of the flow. We found that the decay rate and oscillation frequency divided by $\kappa$, $\sigma/\kappa$ and $\omega/\kappa$, increase in magnitude as $\gamma$ increases from $1$ to $5/3$ and have finite limits as $\gamma\to 1$. In contrast to this, the numerical data suggests that $\omega/\kappa$ diverges polynomially as $\gamma$ approaches the value of $5/3$, which makes sense since the transonic Newtonian flow considered in this article ceases to exist for $\gamma = 5/3$.

We end this article with brief comments on the feasibility of observing or measuring the  quasi-normal acoustic oscillations discussed here. First, let us comment on the frequency scales associated with these oscillations. As we have shown, both the decay rate and the oscillation frequency scale like the surface gravity $\kappa$ of the acoustic black hole. As follows from Ref.~\cite{eCmMoS15}, the surface gravity for a fully relativistic, polytropic flow accreted by a Schwarzschild black hole is
\begin{equation}
\kappa = \sqrt{\frac{9 + (5-3\gamma)(4\xi_c - 3)}{32}}\frac{1}{\xi_c^2} \frac{c}{r_H},
\end{equation}
with $\xi_c = r_c/r_H$ the ratio between the sonic horizon radius and the Schwarzschild radius $r_H = 2GM/c^2$. When the sound speed at infinity $v_\infty$ is much smaller than the speed of light $c$, we have, for $1 < \gamma < 5/3$, $\xi_c \simeq (5-3\gamma)/(8v_\infty^2/c^2)$, and the above expression for $\kappa$ reduces to the Newtonian expression given in Eq.~(\ref{Eq:SurfaceGravity}). For $\gamma = 4/3$ in particular, $\kappa$ is of the order of
$$
\kappa \sim 10^3\left( \frac{M_\odot}{M} \right)\left( \frac{v_\infty}{c} \right)^3 kHz,
$$
with $M_\odot$ the solar mass and $v_\infty$ the sound speed at infinity. Comparing this expression with the corresponding scale for the fundamental quadrupolar quasi-normal frequency of a Schwarzschild black hole (see~\cite{eBvCaS09} and references therein),
$$
f_{BH} \sim 10\left( \frac{M_\odot}{M} \right) kHz,
$$
we see that the fluid oscillations's frequencies are suppressed by a factor of about $100(v_\infty/c)^3$ compared to the quasi-normal frequencies associated with the black hole. For the situation where gas is accreted from the interstellar medium, where $v_\infty \simeq 10km/s$, this factor is extremely small, leading to very large periods for the fluid oscillations even for stellar-mass black holes. The situation improves somehow when considering a fluid flow with adiabatic index $\gamma = 5/3$, in which case $\xi_c \simeq 3/(8v_\infty/c)$ when $v_\infty \ll c$, see for instance~\cite{eCoS15a}. In this case, we obtain
$$
\kappa \simeq \frac{8\sqrt{2}}{3} \frac{c}{r_H}\left( \frac{v_\infty}{c} \right)^2
 \sim 10^2\left( \frac{M_\odot}{M} \right)\left( \frac{v_\infty}{c} \right)^2 kHz,
$$
which is quadratic instead of cubic in $v_\infty/c$; however, for gas accreted from the interstellar medium this is still very small compared to the fundamental quadrupolar quasi-normal frequency of the accreting black hole.

Second, when the backreaction of the fluid on the metric field is taken into account, the quasi-normal acoustic oscillations with angular momentum number $\ell\geq 2$ should produce gravitational radiation which could be modeled, to a good approximation, by considering linear metric perturbations of a Schwarzschild black hole (see Ref.~\cite{LpSS} and references therein for an efficient formalism for describing such perturbations). For a related recent study on gravitational waves triggered by a non-spherical dust cloud which is accreted by a Schwarzschild black hole, see~\cite{jDvGcMdN14}. However, in the scenario discussed in this article it does not seem that the amount of gravitational radiation will be significant since (i) we only consider small departures of a spherically symmetric flow and (ii) because the characteristic frequencies associated with the acoustic perturbations are so much smaller than the characteristic frequencies of the accreting black hole. On the other hand, the situation could change dramatically when considering more realistic flows like those occurring in accretion disks, where the flow has already nonvanishing angular momentum in the unperturbed solution. We intend to generalize our work to such scenarios in future work.

\begin{acknowledgements}
O.S. thanks Luis Lehner for insightful discussions and the Perimeter Institute for Theoretical Physics for hospitality. This research was supported in part by CONACyT Grant No. 238758, by a CIC Grant to Universidad Michoacana and by Perimeter Institute for Theoretical Physics. Research at Perimeter Institute is supported by the Government of Canada through Industry Canada and by the Province of Ontario through the Ministry of Research and Innovation. 
\end{acknowledgements}

\bibliographystyle{spphys}
\bibliography{refs_accretion.bib}

\end{document}